# FPGA-BASED SPILL REGULATION SYSTEM FOR THE MUON DELIVERY RING AT FERMILAB

J. R. Berlioz*, D. Bracey, B. Fellenz, M. A. Ibrahim, V. Nagaslaev,
A. Narayanan, P. Prieto, K. Danison-Fieldhouse, W. Sullivan
Fermi National Accelerator Laboratory†, Batavia, IL, USA

## Abstract

The Muon to Electron Experiment (Mu2e) requires a uniform beam profile from the Muon Delivery Ring to meet their experimental needs. A specialized Spill Regulation System (SRS) has been developed to help achieve consistent spill uniformity. The system is based on a custom-designed carrier board featuring an Arria 10 SoC, capable of executing real-time feedback control. The FPGA processes beam pulses of approximately 200 ns every 1.695 µs, allowing for continuous monitoring of the extracted spill intensity through fast bunch integration. The system directly controls three quadrupole magnets, which work in conjunction with sextupole magnets to achieve third-order resonant extraction. Furthermore, the board interfaces with Fermilab's Accelerator Control Network (ACNET), enabling operators to modify spill regulation settings in real-time via the control network while providing diagnostic waveforms. These waveforms help operators monitor the process and fine-tune the feedback mechanisms. This paper presents an overview of the board's architecture and its initial progress toward regulating beam extraction. This initial version of the regulation system aims to evaluate baseline performance to inform future system improvements.

## INTRODUCTION

Achieving uniform spill intensity during slow resonant extraction is critical for fixed-target experiments like Mu2e, as detector performance and data quality depend strongly on spill uniformity. Factors such as magnet current fluctuations, power supply ripple, and dynamic beam instabilities can degrade spill quality. Thus, effective regulation must quickly counteract these disturbances.

Previous preliminary studies on this challenge explored applying two different regulation schemes [1–3]. One uses RF Knockout (RFKO) striplines, while another uses the fast-ramping tune quadrupoles. In this work we focus on the latter, leaving RFKO-based regulation for future study. The layout for the fast-ramping quadrupole current regulation scheme is illustrated in Fig. 1 and was operationally tested with only extracted beam integration using the first version of the SRS board during a three-month run starting in May 2025.

Three quadrupole magnets regulate the beam tune, facilitating third-order resonant extraction. Increasing quadrupole current progressively moves the beam into the extraction region. A gap monitor provides the measurement of the circulating beam, while a wall current monitor (WCM) in the extraction line measures the extracted beam. An example of the extracted spill structure is shown in Fig. 2, which displays a 20 ms segment recorded without feedback loop. Because both monitors are AC-coupled systems, the signal shows a shifting baseline. The lower trace highlights the microbunch structure, which occurs every 1.695 µs with a pulse width of approximately 200 ns. These beam measurement serve as the primary inputs to the regulation system. The extracted spill intensity from the WCM, in particular, is processed by the SRS board to generate corrective actions on the quadrupole magnets.

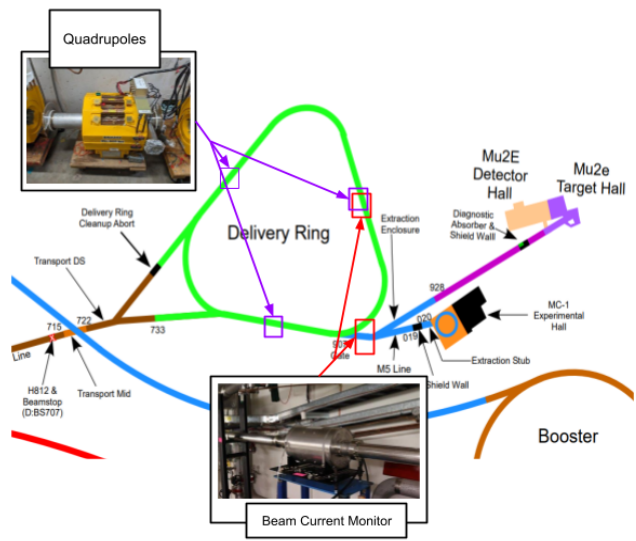

Figure 1: Layout of the three tune quadrupoles in the Delivery Ring. Beam current monitors provide measurements of both the circulating and extracted beam for the feedback loops.

## HARDWARE SETUP

The Mu2e SRS board used is shown in Fig. 3. It features sixteen 14-bit ADC channels at 125 MSPS, sixteen precision outputs from four DACs, and four high-speed outputs from two dual DACs. The main board is a ReflexCES Arria 10 SoM Turbo, which integrates an Intel Arria 10 GX FPGA with embedded dual-core ARM Cortex-A9 processors. This system-on-module provides high-performance logic resources and embedded processing capabilities, enabling both low-latency control loops and higher-level supervisory tasks to run on the same platform.

___________
* jberlioz@fnal.gov
† This manuscript has been authored by FermiForward Discovery Group, LLC under Contract No. 89243024CSC000002 with the U.S. Department of Energy, Office of Science, Office of High Energy Physics.

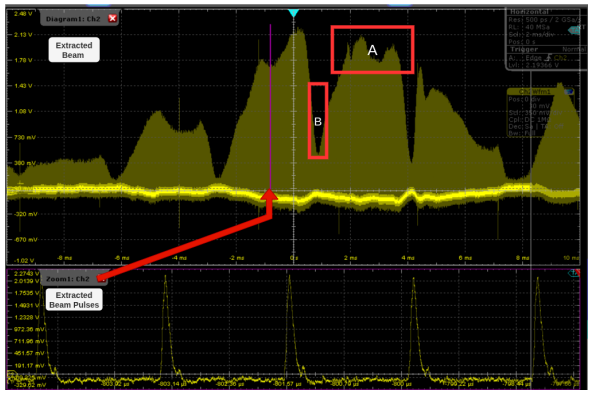

Figure 2: Oscilloscope capture of the extracted spill showing a high-intensity peak (A) and a low-intensity notch (B) linked to the 300 Hz modulation. Bunch structure capture is shown on the bottom window.

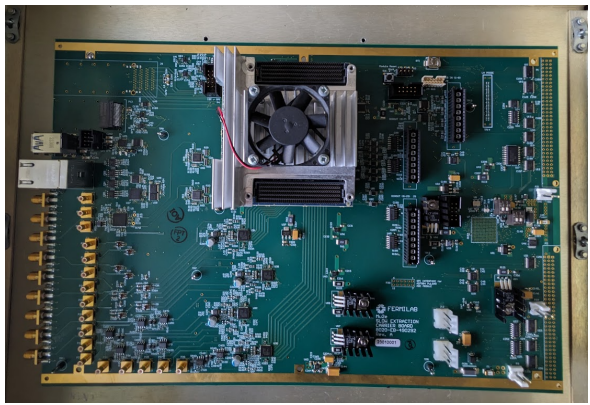

Figure 3: Mu2e Spill Regulation System board.

Data exchange between FPGA and ARM occurs via Avalon/AXI bridges, facilitating control through Linux file operations. With this setup, the memory space on the FPGA board is mapped to an external Redis database, which enables communication with ACNET, crucial for updating waveforms and tuning control algorithms.

## SYSTEM ARCHITECTURE

Figure 4 illustrates the overall system. The regulation system consists of two primary components: a *spill playback generator*, updated periodically (minutes to hours) after multiple spills, and a *fast regulation controller*, operating at 10 kHz intervals. The green boxes in Fig. 4 show the intended location of the regulation systems. Notice that the slow regulation loop happens outside the board and is loaded during the boards idle periods through the spill playback generator.

The spill playback generator delivers the initial reference spill waveforms at 10 kHz with high resolution. Profiles can be uploaded through Redis/ACNET interfaces or directly from the ARM processor via C++, and are stored in spill buffers synchronized across the system. An internal, configurable timer ensures deterministic playback from on-board memory with fine granularity.

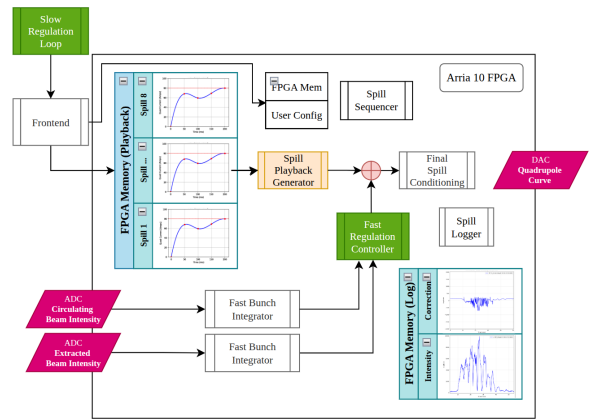

Figure 4: SRS Board architecture for quadrupole extraction scheme.

The fast feedback controller performs rapid real-time corrections during the spill. The reference spill waveforms are dynamically refined by the fast feedback controller. Currently, a PID controller is implemented; however, additional harmonic correction may be necessary. Additional work is being done to create a more robust non-linear controller using an ML model, which is being developed in parallel with the classical controller. A final conditioning stage includes a safe ramp-down mechanism activated by either abort events or an internal timeout, as well as a slew rate limiter as an additional safety mechanism.

The fast bunch integration modules feed the beam intensities to the fast feedback controller and buffer data for external analysis. The fast bunch integrator works by using an external RF trigger. Once the trigger is activated, a static window integrates all ADC samples within a specified window after a set delay, as shown in Fig. 5.

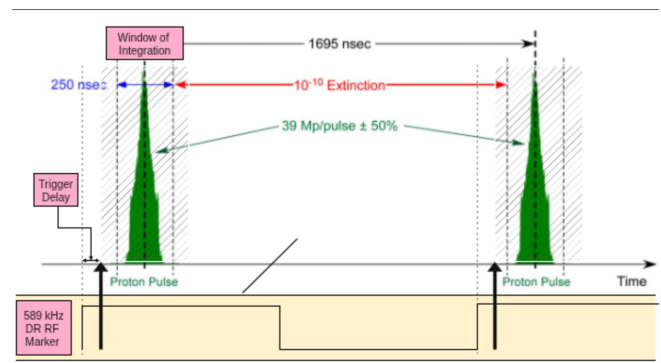

Figure 5: Fast bunch integration performed with a fixed, trigger-aligned window.

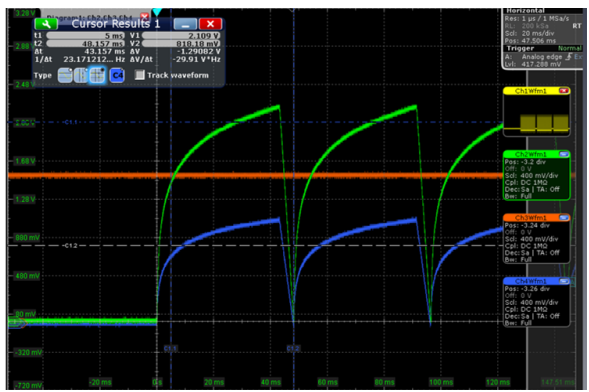

Figure 6: Oscilloscope showing the quadrupole current (green trace) tracking the board's reference signal (blue trace).

## RESULTS

This section presents the results from the Mu2e operational run conducted between May and July 2025, highlighting the performance and evaluation of the spill regulation system under beam conditions.

### Waveform Playback Generator

During this test period, the system was operated without the slow regulation algorithm to evaluate the end-to-end signal path. Arbitrary waveforms were successfully delivered from `acsys-python` to the quadrupoles at 10 kHz. Figure 6 illustrates the quadrupole power supply's load current tap tracking an arbitrary logarithmic curve injected through the board via `acsys-python`.

Looking ahead, spill profiles will be derived from the logged intensity and error spill waveforms observed over multiple spills.

### Fast Bunch Integration

The SRS board also captured and integrated live data synchronized to both the spill events and RF markers. Figure 7 shows the resulting integration after baseline correction, as displayed in the ACNET Array Plotter. To validate the integration procedure, we compared it to data captured by an oscilloscope, for which an intensity measurement was obtained using Python, and by replaying the full integrated signal—prior to moving-average smoothing and 10 kHz PID decimation—through a DAC channel for direct comparison with the raw WCM waveform. Calibration to absolute proton counts will be carried out once the WCM calibration is complete; until then, results are expressed in raw ADC integration units.

### Fast Feedback Controller

In this run, a PID controller was evaluated as the fast feedback regulator for a limited duration. Testing revealed that the loop was sensitive to strong 300 Hz modulations present in the spill. These periodic notches were sufficient to induce oscillations, reducing the controller's effectiveness.

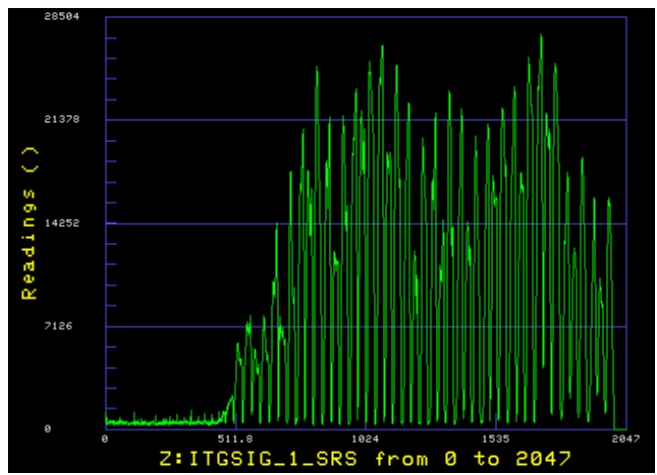

Figure 7: Raw integrated beam pulses from a wall current monitor, displayed using Fermilab's ACNET Array Plotter.

Characterizing these effects provides useful insight for improving future iterations of the regulation system.

## CONCLUSIONS

The first version of the Mu2e Spill Regulation System has been successfully integrated with the Delivery Ring and tested under beam conditions. We demonstrated that the waveform playback generator can deliver high-resolution reference signals, and that the fast bunch integrator provides stable and repeatable intensity measurements suitable for feedback control. The PID-based feedback loop revealed important challenges, particularly the sensitivity to 300 Hz notches, which highlights the complexity of regulating spill uniformity in practice.

These initial results establish a performance baseline and validate the core features of the system. More importantly, they provide critical insights that will guide the next iteration of development, including improved filtering strategies, refined control algorithms, and integration of adaptive techniques. For instance, a median filter will be explored as an alternative to the current moving average, offering potentially faster responsiveness with fewer samples. Additionally, harmonic correction will be incorporated into the fast regulation controller to mitigate the strong 300 Hz modulations observed during initial testing. In parallel, the fan-c project is investigating machine learning approaches to handle nonlinearities in the disturbances, which could replace our classical control techniques [4–6]. By systematically addressing these challenges, the SRS will move closer to delivering the uniform and stable spills required for the high-precision physics goals of the Mu2e experiment.